\documentstyle[12pt]{article}

\catcode`\@=11

\@addtoreset{equation}{section}
\def\theequation{\arabic{section}.\arabic{equation}}

\def\@normalsize{\@setsize\normalsize{15pt}\xiipt\@xiipt
\abovedisplayskip 14pt plus3pt minus3pt%
\belowdisplayskip \abovedisplayskip
\abovedisplayshortskip  \z@ plus3pt%
\belowdisplayshortskip  7pt plus3.5pt minus0pt}

\def\small{\@setsize\small{13.6pt}\xipt\@xipt
\abovedisplayskip 13pt plus3pt minus3pt%
\belowdisplayskip \abovedisplayskip
\abovedisplayshortskip  \z@ plus3pt%
\belowdisplayshortskip  7pt plus3.5pt minus0pt
\def\@listi{\parsep 4.5pt plus 2pt minus 1pt
            \itemsep \parsep
            \topsep 9pt plus 3pt minus 3pt}}

\def\underline#1{\relax\ifmmode\@@underline#1\else
        $\@@underline{\hbox{#1}}$\relax\fi}
\@twosidetrue
\relax

\catcode`@=12

\catcode`\@=11

\def\section{\@startsection{section}{1}{\z@}{3.5ex plus 1ex minus
   .2ex}{2.3ex plus .2ex}{\large\bf}}

\def\thesection{\Roman{section}.}

\def\appendix{\setcounter{section}{0}
        \def\thesection{APPENDIX }
        \def\theequation{\Alph{section}.\arabic{equation}}}
\def\ps@headings{\def\@oddfoot{}\def\@evenfoot{}
\def\@oddhead{\hbox{}\hfill
        \makebox[.5\textwidth]{\raggedright\ignorespaces --\thepage{}--
        \hfill {}}}
\def\@oddhead{\hbox{}\hfill --\thepage{}-- \hfill
        {}}
\def\@evenhead{\@oddhead}
\def\subsectionmark##1{\markboth{##1}{}}
}

\ps@headings

\catcode`\@=12

\relax

\def\figcap{\section*{Figure Captions\markboth
        {FIGURECAPTIONS}{FIGURECAPTIONS}}\list
        {Fig. \arabic{enumi}:\hfill}{\settowidth\labelwidth{Fig. 999:}
        \leftmargin\labelwidth
        \advance\leftmargin\labelsep\usecounter{enumi}}}
 \relax
\def\tablecap{\section*{Table Captions\markboth
        {TABLECAPTIONS}{TABLECAPTIONS}}\list
        {Table \arabic{enumi}:\hfill}{\settowidth\labelwidth{Table 999:}
        \leftmargin\labelwidth
        \advance\leftmargin\labelsep\usecounter{enumi}}}
 \relax
\def\reflist{\section*{References\markboth
        {REFLIST}{REFLIST}}\list
        {[\arabic{enumi}]\hfill}{\settowidth\labelwidth{[999]}
        \leftmargin\labelwidth
        \advance\leftmargin\labelsep\usecounter{enumi}}}
 \relax

\catcode`\@=11

\def\ps@headings{\def\@oddfoot{}\def\@evenfoot{}
\def\@oddhead{\hbox{}\hfill
        \makebox[.5\textwidth]{\raggedright\ignorespaces --\thepage{}--
        \hfill {}}}
\def\@evenhead{\@oddhead}
\def\subsectionmark##1{\markboth{##1}{}}
}

\ps@headings

\relax

\newskip\humongous \humongous=0pt plus 1000pt minus 1000pt

\newif\ifdtup

\def\Im{\mathop{\rm Im}}
\def\Re{\mathop{\rm Re}}

\def\beq{\begin{equation}}
\def\eeq{\end{equation}}

\def\beqn{\begin{eqnarray}}
\def\eeqn{\end{eqnarray}}
\relax

\def\G2{{\; \rm GeV/}c^2}
\def\G{\; \rm GeV}

\def\dotx{\dotx{\dot\overline{x}}}

\relax

\begin{document}

\begin{titlepage}
\nopagebreak
\begin{flushright}

        {\normalsize     OU-HET-182  \\
                         September,~1993\\}

\end{flushright}

\vfill
\begin{center}
{\large \bf           Sine-Gordon Theory with \\
   Higher Spin $N=2$ Supersymmetry\\
            and the Massless Limit }\footnote{This work is supported in part by
  Grant-in-Aid for  Scientific Research
  $(05640347)$  from the Ministry of Education, Japan.}

%\vfill
\vfill

        {\bf H.~Itoyama} and {\bf T. Oota}\\

        Department of Physics,  Faculty of Science, \\
        Osaka University,  Toyonaka, Osaka 560, Japan \\

\end{center}
\vfill

\begin{abstract}
  The Sine-Gordon theory at
 $\frac{\beta^{2}}{8\pi} = \frac{2}{(2n+3)},\;
 n= 1,2,3 \cdots $
 has a higher spin generalization
 of the $N=2$ supersymmetry  with the central terms which
       arises from the affine quantum group  $U_{q}( \hat{s \ell} (2))$.
  Observing that  the  algebraic determination of $S$ matrices $( \approx
  {\rm  quantum~
 integrability  })$
  requires the saturation of  the  generalized Bogomolny bound,
  we construct  a variant
of the Sine-Gordon theory  at this value of the coupling   in the
 framework of  $S$ matrix theory.
   The spectrum  consists of a doublet of
  fractionally charged solitons as well as that of  anti-solitons
   in addition to the ordinary breathers.
The construction   demonstrates  the existence of the theory other than
 the one by the  truncation  to the
 breathers considered by Smirnov.
   The allowed values for the fractional part of  the fermion
 number is also determined.
 The central charge in the massless limit is  found to be
$c= 1$  from the TBA calculation for  nondiagonal S matrices.
  The attendant $c=1$ conformal field theory
  is the gaussian model with  ${\bf Z_{2}}$ graded chiral algebra
  at the radius parameter $r= \sqrt{2n+3}$.
 In the course of the calculation, we find  $4n+2$  zero
 modes from  the (anti-)soliton distributions.

\end{abstract}
\vfill
\end{titlepage}

\section{Introduction}
\label{Introduction}

   The Sine-Gordon theory has been a prototype of
  relativistic  continuum field theories that are integrable
 in $1+1$ dimensions.
    The mass spectra and $S$ matrices  as well
 as the relationship with the massive Thirring model
  have been well-studied  for a long time \cite{SG,ZZ}.
  In recent years, attention has been paid to identify the universality class
  which the model belongs to,  for a variety of   values
 of the coupling constant.
 This is not a trivial question as the model at  special values
 of the coupling constant
 permits  a self-consistent truncation, which can be revealed by the
  consistency of the bootstrap framework or by the structure of the
  vacuum. This  point becomes evident in view of the perturbation of
 the conformal field theory at  the fixed point \cite{REV}.
  Formulating thermodynamics from the physical S matrices  \cite{ALZ},
 often referred to as
 thermodynamic Bethe ansatz (TBA)\cite{YangYang},
 provides an efficient method  to reveal  ground state properties
 of the model.  In some cases, it is sufficient to identify
  the attendant conformal field theory.

  One series of special points considered so far  is
\beqn
\label{eq:Smir}
  \frac{ \beta^{2}}{ 8\pi}
 =  \frac{2}{2n+3}\;\;\;. \;\;\; n= 1,2,3, \cdots
\eeqn
   This lies in the  regime in which the Thirring coupling is attractive.
  It was noted \cite{Smirnov}, on the basis of the structure of
 the form factors, that  keeping breathers alone is a self-consistent
 truncation of the model. In  \cite{Smirnov},
  the massless limit of this series of models with truncation
 has been identified with
 the nonunitary  minimal conformal field theory indexed by
  $(2,2n+3)$.
   The  TBA calculation has confirmed this \cite{KM}.
  Another series of points where the TBA calculation  has been
 done \cite{IM,ALZSOS} includes
\beqn
\label{eq:min}
   \frac{\beta^{2}}{8\pi} = 1- \frac{1}{\ell +1}\;\;
. \;\;\; \ell
  = 2,3,4 \cdots \;\;\;,
\eeqn
 lying in the repulsive regime.  ( See \cite{ChAh} for TBA  at  other series of
points.)
  Both the calculation \cite{IM}
 based on the structure of the vacuum considered from the microscopic Bethe
 Ansatz diagonalization \cite{Korepin2}
 and the one based on the ``RSOS'' scattering
 matrices proposed \cite{ALZSOS} have led to the same set of the algebraic
equations.
   The central charge in the massless limit is  found to be
   that of  the  unitary minimal model
  $c=  1 - \frac{6}{\ell(\ell + 1)}$ \cite{IM,ALZSOS}.
The neutral excitations (or distributions)
 play a prominent role here
 \footnote{ Similar features have been seen in the microscopic calculation
 of ref.~\cite{BR} for the critical  RSOS spin chain.}.

{}From the representation-theoretical point of view,
 it is well-known that the soliton-(anti-)soliton $S$ matrices of
  the Sine-Gordon are naturally
  regarded as the  trigonometric $\check{R}(v)$  matrices in the vector
representation
 which have the quantized universal enveloping algebra \cite{D,Jimbo}
  ( or the affine quantum group) $ U_{q}(\hat{s \ell}(2))$ \cite{KS}
 as a commutant ( or symmetry):
 \beqn
\label{eq:rdelta}
  \left[ \check{R}(v), ~  \pi \otimes \pi\left( \Delta \left(
  \phi( X) \right) \right) \right] = 0 \;\;\;, \;\;\;
   X \in  U_{q}(\hat{s \ell}(2)) \;\;\;.
 \eeqn
  Here $\pi$ denotes the representation and $\Delta$ is the co-product.  In
order
  to find a finite dimensional representation, $e_{0}$ and $f_{0}$ have been
 represented through the restriction  $ \phi$ \cite{Jimbo}
 of   $U_{q}(\hat{s \ell}(2))$
  onto the non-affine part  $U_{q}(s \ell(2))$.  ( See Appendix A.)
In what follows, we will employ eq.~(\ref{eq:rdelta}) as a definition
 of the word ``Sine-Gordon''.
  The  theory, therefore, has   $U_{q}(\hat{s \ell}(2))$
  at
 \beqn
   q=  e^{- \frac{8 \pi^{2}i}{\beta^{2} }  } \;\;\;
 \eeqn
 as  genetic
 on-shell symmetry of  $S$ matrices.
  It has been noted that, at $q^{2} = -1$,  some of the defining relations
 of $ U_{q}(\hat{s \ell}(2))$ can be converted into
 anti-commutators \cite{PAS,BBL}.
  In this vein,   Smirnov's truncation
 at eq.~(\ref{eq:Smir}), which
 yields $q^{2}=-1$, is the case in which the symmetry
    $U_{q}(\hat{s \ell}(2))$  is nonlocally
  realized. In more physical terms, it is
 the one in which  ``charge carriers'' are confined by a  dynamical reason.

    An interesting variant has been considered  by Fendley and Intrilligator
  \cite{FI}
  at  $ \frac{ \beta^{2}}{ 8\pi} = \frac{2}{3}.$
   They   propose   $S$ matrix theory   where the spectrum consists
 of a doublet of the solitons and that of the anti-solitons.
   This is distinct from the original Sine-Gordon thoery. Each of the two
 doublets provides a representation of the $N=2$ supersymmetry algebra
 with the central terms \cite{WO}, which is possible only when
  the  Bogomolny bound is saturated. The solitons and the anti-solitons
 carry  non-integer fermion numbers
  \cite{JR}.   This theory
 is interpreted  as the one in which the symmetry  $ U_{q}(\hat{s \ell}(2))$
 is locally  realized by the  doublet of solitons and that of anti-solitons.
  The central charge $c$ in the massless limit is found to be
 $c=1$ \cite{FI}.    Note
 that  this point   is the intersection  ($n=0$) of  the two series
   listed  in eqs.~(\ref{eq:Smir}) and (\ref{eq:min}),
 where the theory considered  in \cite{IM,ALZSOS}
  predicts  $c=0$.
 The existence of the  $c=1$  $S$ matrix theory of \cite{FI}
    at  $\frac{ \beta^{2}}{ 8\pi} = \frac{2}{3}$
  suggests that  there be two distinct ($S$ matrix) theories for
 every point of the series eq.~(\ref{eq:Smir}): the one is that
   of \cite{Smirnov} where   $ U_{q}(\hat{s \ell}(2))$ is nonlocally
 realized  and the other is the possibility  in which
 $ U_{q}(\hat{s \ell}(2))$ is locally realized
 and which has not been considered before.

 In what follows, we  propose  a variant of the Sine-Gordon theory
at   $\frac{\beta^{2}}{8\pi}=\frac{2}{2n+3}$
  as  $S$-matrix theory,  filling up this missing possibility.
  The  $ U_{q}(\hat{s \ell}(2))$  is locally realized by
 a doublet of solitons  as well as that of anti-solitons
  whereas the breathers  ( neutral particles)
 are considered as a nonlocal realization of
  $ U_{q}(\hat{s \ell}(2))$.
  The argument from perturbed conformal field theory tells us that
  the Lorentz spin of the anticommuting charges  is
 $  \pm \frac{1}{\gamma}= \pm \left(  \frac{8\pi}{\beta^{2}} -1  \right)
 =  \pm \left( n+1/2  \right)$ \cite{BBL}. The algebra is,
 therefore, regarded as a higher spin generalization of the $N=2$
  supersymmetry algebra with central terms.
  ( Examples of conserved charges which form a higher spin analog of
supersymmetry have been given in \cite{Witten,Sch}.)
    The two doublets again form representations where   the  Bogomolny
 bound is saturated.

In the next section, we first  derive the higher spin generalization
 of $ N=2$ supersymmetry with the central charge from
   $ U_{q}(\hat{s \ell}(2))$ at $q = \pm i$.
  We note that  the homomorphism $\phi :U_{q}(\hat{s \ell}(2))
     \rightarrow U_{q}(s \ell(2)) $   quoted in Appendix A
  is equivalent to the statement  that the Bogomolny bound
   is saturated.
   We construct the soliton-antisoliton $S$ matrices by demanding
  the algebra to be  on-shell symmetry.
  This procedure also determines
 the allowed values of the fractional part of
 the fermion number.

  In section $3$, we diagonalize
 the TBA  for the nondiagonal soliton-antisoliton $S$-matrices  by utilizing
 the quantum inverse scattering method
 ( or algebraic Bethe ansatz) \cite{QIM}\footnote{A part
 of  section $3$  is a direct
 generalization of the calculation carried out
 in ref.~\cite{FI} for the $n=0$ case. For the sake of the clarity
 of this paper, we give some detail of our calculation.}.
 We find $4n+2$ fermionic zero modes around the (anti-)soliton  distributions
 after the diagonalization.
In section $4$, we present  our calculation of the
   the central charge in the massless limit. We find
\beqn
   c= 1 \;\;,\;\;\;\;\;\; {\rm for}\;\; n= 1,2,3 \cdots
\eeqn
 In section $5$,  the  $c=1$ conformal field theory
  attendant with
  our model is identified to be the gaussian model having
 the radius
 parameter $r= \sqrt{2n+3}$ and a ${\bf Z_{2}}$ graded chiral algebra.

  Appendix A $\sim$ C  succinctly  summarize
     $ U_{q}(\hat{s \ell}(2))$,  TBA for diagonal scattering and
  some formulas related to  the Rogers dilogarithm  respectively.
   In Appendix D, we present  a  calculation  with  imaginary
 chemical potentials.

\section{  Construction of $ S $ Matrices from    $ U_{q}(\hat{s \ell}(2))$
   at $q =  \pm i$  }

 Let us first  derive the higher spin generalization
 of  $ N=2$ supersymmetry algebra with the central terms from
   $ U_{q}(\hat{s \ell}(2))$ at $q = \pm i$.
  From the generators of  $ U_{q}(\hat{s \ell}(2))$  introduced in the Appendix
$A$,
 we define
\beqn
\label{eq:defn}
   {\cal F} &\equiv&  \frac{  \phi (h_{1}) }{2}\;\;\;,\;\;\;
   Q^{+} \equiv M^{n+1/2}  \phi ( e_{1}) q^{{\cal F} }\;\;\; ,\;\;\;
   Q^{-} \equiv   M^{n+1/2}  \phi ( e_{0})
 q^{-{\cal F} }\;\;\; ,\; \;\;
  \nonumber \\
   \overline{Q}^{+} &\equiv&
   M^{n+1/2} \phi ( f_{1}) q^{{\cal F}}\;\;\;,\;\;\;
 \overline{Q}^{-} \equiv   M^{n+1/2}
   \phi ( f_{0}) q^{-{\cal F} } \;\;\;.
\eeqn
  Here  we have introduced a mass scale $M$  and $\phi$ is the homomorphism
defined in eq.~(\ref{eq:hom}).
 Eqs.~(\ref{eq:hef}),(\ref{eq:qse}) and (\ref{eq:qsf})  in the Appendix $A$
become
\beqn
\label{eq:SUSY}
  &\left[{\cal F}, Q^{\pm} \right] = \pm Q^{\pm} \;\;,\;\;
   &\left[ {\cal F}, \overline{Q}^{\pm}  \right] = \mp
    \overline{Q}^{\pm}  \;\;\;,\\
   &\left \{ Q^{+}, \overline{Q}^{+} \right \} = {\cal Z}^{(+)} \;\;, \;\;
     &\left \{ Q^{-}, \overline{Q}^{-} \right \} =
 {\cal Z}^{(-)} \;\;, \;\;
    \left \{ Q^{+}, \overline{Q}^{-} \right \} = 0 \;\;,\;\;
     \left \{ Q^{-}, \overline{Q}^{+} \right \} = 0  \;\;,\;\;  \nonumber
\eeqn
  where
\beqn
\label{eq:Z}
 {\cal Z}^{(\pm)} \equiv \frac{  M^{2n+1}}{2}
 \left( 1- q^{\pm 4 {\cal F} }
 \right)
\eeqn
 commutes with   ${\cal F}, Q^{\pm},
  \overline{Q}^{\pm} $
 and in fact a center of
 the algebra generated by these.   In  the second  line
 of eq.~(\ref{eq:SUSY}),
 we have used $q = e^{-\pi i  (n+ 3/2)}$.
  Eq.~(\ref{eq:Z}) tells us that the center can be represented
 as a nonvanishing number only when the fermion number
  of the state is fractional.

Define
 $ {\cal P} \equiv \{ Q^{+}, Q^{-} \},\;
  \overline{ {\cal P}} \equiv \{ \overline{Q}^{+},\overline{Q}^{-}  \}.\;$
  The $q$ analog  of the
 Serre relations ( eqs.~(\ref{eq:qse}),~(\ref{eq:qsf}) )
 reads
\beqn
\label{eq:serre}
   \left[ {\cal P}, \left(Q^{\pm}\right)^{2} \right] = 0  \;,\;\;\;
     \left[ \overline{ {\cal P}}, \left(\overline{ Q}^{\pm} \right)^{2}
 \right] = 0\;\;\;
\eeqn
    for $q = e^{-\pi i  (n+ 3/2)}$.  It is known  that one can set
\beqn
\label{eq:q2zero}
     \left(Q^{\pm}\right)^{2}=0 \;\;, \;\;\;
  \left(\overline{ Q}^{\pm} \right)^{2}  =0 \;\;\;
\eeqn
  consistently \cite{Lustig,LV}.
We also find,    for $q = e^{-\pi i  (n+ 3/2)}$,
\beqn
\label{eq:comm}
  \left[ {\cal P}, Q^{\pm } \right] =  - \left[  (Q^{\pm })^{2},
 Q^{\mp }  \right]  = 0 \;\; ,\;\;\;
     \left[  {\cal P},\overline{Q}^{\pm } \right] = 0\;\; ,  \nonumber  \\
  \left[ \overline{ {\cal P}}, \overline{Q}^{\pm } \right] =
  - \left[   (\overline{Q}^{\pm })^{2} ,   \overline{Q}^{\mp } \right]
  = 0 \;\; , \;\;\;
     \left[ \overline{ {\cal P}}, Q^{\pm } \right] = 0 \;\; ,
\eeqn
 from a simple calculation.
  The algebra defined by eqs.~(\ref{eq:SUSY}), (\ref{eq:q2zero})
 and (\ref{eq:comm})
 is the higher spin  $N=2$ supersymmetry algebra with the central terms.
 The co-product (eq.~(\ref{eq:cop})) is translated into
  \beqn
     \Delta (Q^{\pm})=Q^{\pm}\otimes 1+  q^{\pm 2{\cal F}}
 \otimes
  Q^{\pm} \;\;\;,\;\;\;
     \Delta (\overline{Q}^{\pm})= \overline{Q}^{\pm}\otimes
 1+  q^{\pm 2{\cal F}} \otimes \overline{Q}^{\pm} \;\;\;.
  \eeqn

  We consider  finite dimensional representation of the algebra
   in which  ${\cal P}$ and $\overline{\cal P}$ are diagonal.
 Writing the eigenvalue of ${\cal P}$ and that of $\overline{\cal P}$  as
 $m^{2n+1} e^{(2n+1) v}$ and  $m^{2n+1} e^{-(2n+1) v}$
 respectively, we  label
  a generic state by $m$ and $v$, which  are considered as
 the mass and the rapidity of  the particle,
  and by the fermion number $f$.
  As in the case of the ordinary $N=2$ supersymmetry,
 the Bogomolny bound can be derived
 \cite{WO} by considering  the positive
 semi-definiteness  of $\{ A^{\dagger}, A \}$ with
\beqn
 A \equiv
 Q^{+} \left(\overline{ {\cal P}} {\cal Z}^{(-)} \right)^{1/2}
     - \overline{Q}^{-} \left({\cal P} {\cal Z}^{(+)} \right)^{1/2}
\eeqn
: this
 gives  $ M^{4n+2} \sin^{2}( (2n + 3)\pi f)
 \leq  m^{4n+2}  $.
  The two-dimensional
 representation is possible only when
 the bound gets saturated and
 the central  terms are nonvanishing.  The latter condition
 leads to the  non-integer fermion number.

    We note that the homomorphism   $\phi$
 given in eq.~(\ref{eq:hom})
 is equivalent to   $A=0$ and, therefore, implies
  the saturation of the Bogomolny bound.
 In fact, by identifying
\beqn
  \lambda =  i e^{ (2n+1) v} \;\;\;,
\eeqn
  we see that eq.~(\ref{eq:hom}) via
 eq.~(\ref{eq:defn}) translates into $A=0$ evaluated
  at the one-particle states.  Quantum
integrability,  which permits us to construct $S$ matrices
  algebraically, demands that the bound  be saturated.
 The old wisdom is now found to be a reality.

  We are thus led to consider the two dimensional representation
 with non-integer fermion number.
  Following ref. \cite{FI},
 we postulate that the soliton and the anti-soliton individually
 form  doublets in contrast to the original Sine-Gordon case.
  The rationale for the two doublets comes from the
   semi-classical  description of the corresponding
  Landau-Ginzburg Lagrangian \cite{FI,LG} and is closely  related
 to the chiral algebraic structure of the theory as we will see.
The soliton doublet and the anti-soliton doublet are denoted  respectively
 by $\left(  \begin{array}{c} u \\ d  \end{array}
 \right)$ and $\left( \begin{array}{c}\bar{d} \\ \bar{u}
 \end{array}  \right)$.
The fermion number of the  soliton doublet  is denoted by
 $\left(  \begin{array}{c} f \\ f-1   \end{array} \right)$
 and that of the anti-soliton doublet  by
  $\left(  \begin{array}{c} 1-f \\  -f   \end{array} \right)$.
   In the case in which more than one doublet
  are present in the theory, the charge
 conjugation symmetry alone  does not  force the fermion number
 to be  $ \pm \frac{1}{2}$ \cite{SSH}.

The action of the charges  on  the one-soliton states
 must be consistent with eq.~(\ref{eq:SUSY}).  Determining
 the phases  allowed, we find
  \begin{eqnarray}
\label{eq:action1}
     Q^{+}\mid d(v)>&=&(me^{v})^{n+\frac{1}{2}}\mid u(v)>,
	                                                    \;\;\;
     Q^{-}\mid u(v)> ~=~(me^{v})^{n+\frac{1}{2}}\mid d(v)>,
	                                                  \\
     \overline{Q}^{+}\mid u(v)>&=&
                e^{i\alpha}(me^{-v})^{n+\frac{1}{2}}\mid d(v)>,
		                                 \;\;\;
     \overline{Q}^{-}\mid d(v)>  ~=~
              e^{-i\alpha}(me^{-v})^{n+\frac{1}{2}}\mid u(v)> \;.
            \nonumber
  \end{eqnarray}
Here  $e^{i\alpha}=  i e^{- (2n+3)\pi i f} $.
   The other action of the charges on the one-soliton states
 vanishes.
  The equations for  the nonvanishing action of the charges on
 the one anti-soliton states  are the same as eq.~(\ref{eq:action1}).

We write a generic two-soliton state as
     $ \mid A(v_{1})B(v_{2})>=\mid A(v_{1})>
        \otimes \mid B(v_{2})>$.
 In accordance with the co-product (eq.~(\ref{eq:cop})),
 the action of $Q^{\pm}$ on this tensor product state  is
  \beq
     Q^{\pm}\mid AB>=(Q^{\pm}\mid A>)\otimes \mid B>
                             +(q^{\pm 2{\cal F}}\mid A>)\otimes
 (Q^{\pm}\mid B>)  \;\;\;.
  \eeq
  The same holds true for $\overline{Q}^{\pm}$.

 We now turn to consider the elastic scattering among
  the  solitons, the antisolitons and the breathers.
  As is discussed in the introduction,  we will construct a variant
 of the Sine-Gordon  $S$ matrices at  $\frac{\beta^{2}}{8\pi} =
 \frac{2}{(2n+3)}$. The breather-breather scattering and the
 breather-(anti-)soliton  scattering
 are taken from those of the original Sine-Gordon $S$ matrices.
  We determine the soliton-antisoliton $S$ matrices
  by demanding that the  algebra
 (eqs.~(\ref{eq:SUSY}) and (\ref{eq:comm}))
 be on-shell symmetry {\it i.e.} the symmetry
 of the $ S$ matrices.

 Let the S-matrix of the soliton-antisoliton scattering be
 \beq
\label{eq:Smatrix}
     \bordermatrix{          & u(v_{2}) \bar{u}(v_{1})
 & d(v_{2})   \bar{d}(v_{1}) \cr
                    u(v_{1})\bar{u}(v_{2}) & c(v_{1}- v_{2})
  & b(v_{1} - v_{2})        \cr
                    d(v_{1})\bar{d}(v_{2}) & b(v_{1} - v_{2})
    & c(v_{1}-v_{2})        \cr
		  		                                    },\ \ \
     \bordermatrix{          & u(v_{2})  \bar{d}(v_{1}) &
     d(v_{2})  \bar{u}(v_{1}) \cr
                    u(v_{1})\bar{d}(v_{2}) & a(v_{1}- v_{2})        & 0
   \cr
                    d(v_{1})\bar{u}(v_{2}) & 0
 & a(v_{1}- v_{2})
    \cr     }.
  \eeq
  Let us demand that the charges commute with the S-matrix.
  We find that    $f$ must satisfy
$q^{-2(2f-1)} = 1$.  This determines the   allowed set of
  values  $\{ \{  f _{\ell} \} \}$
 for the fractional part of the fermion number:
\beqn
  f_{\ell} = \frac{2(n+1-\ell) +1}{4n+6}\;\;,\;\;\;  \ell \in {\bf Z}
\;\;, \;\;\;
  - (n+\frac{3}{2}) < \ell < n+ \frac{3}{2} \;\;\;.
\eeqn
 We also
 find that $a(v),\ b(v),$ and $ c(v)$ must be of the form
  \beq
   \label{eq:Smatrix2}
     a(v)=  Z(v) \cosh (n+\frac{1}{2})v,\ \
     b(v)= i(-1)^{n-\ell+1}   Z(v) \sinh (n+\frac{1}{2})v,\ \
     c(v)=  Z(v) \;\;\;.
  \eeq

The crossing symmetry and the unitarity
  respectively imply
\beq
 Z(i\pi-v)=Z(v) \;\;\;.
\eeq
and
\beq
Z(v)Z(-v)=  \frac{1}
{\cosh^{2}(n+\frac{1}{2})v} \;\;\;.
\eeq
 These two requirements together with the requirement  of no additional
 bound state completely
determine the ``minimal'' $Z$-factor which has no pole in the
physical strip except for a sign:
  \begin{eqnarray}
    &~& Z_{min}(v)   \nonumber \\
     &=& \frac{(-1)^{n}}{\cosh(n+\frac{1}{2})v}
	      \prod_{j=1}^{\infty}
            \frac{\Gamma^{2}(-\frac{(n+1/2)v}{\pi i}+(2n+1)j-n)
                    \Gamma(\frac{(n+1/2)v}{\pi i}+(2n+1)j+\frac{1}{2})}
                 {\Gamma^{2}(\frac{(n+1/2)v}{\pi i}+(2n+1)j-n)
                    \Gamma(-\frac{(n+1/2)v}{\pi i}+(2n+1)j+\frac{1}{2})}
	                                        \nonumber \\
         &~&  \ \ \ \ \ \ \times
        \frac{\Gamma(\frac{(n+1/2)v}{\pi i}+(2n+1)j-2n-\frac{1}{2})}
             {\Gamma(-\frac{(n+1/2)v}{\pi i}+(2n+1)j-2n-\frac{1}{2})}
                                                        \nonumber \\
     &=&\frac{(-1)^{n}}{\cosh(n+\frac{1}{2})v}
        \exp\left(   i\int_{0}^{\infty}\frac{dt}{t}
              \frac{\sin((n+\frac{1}{2})v t/\pi)\sinh((n+\frac{1}{2})t/2)}
                   {\sinh t/2\cosh(n+\frac{1}{2})t/2}
		    \right).
  \end{eqnarray}
Here we have used the Malmsten's formula
  \beq
     \log\Gamma(z)=\int_{0}^{\infty}\frac{dt}{t}e^{-t}
       \left[  \frac{e^{(1-z)t}-1}{1-e^{-t}}+z-1
  \right] \ \ \ ({\Re}~z>0)\;\;\;.
  \eeq

The value of coupling $\frac{\beta^{2}}{8\pi}=\frac{2}{2n+3}$ is
in the attractive regime (apart from $n=0$ which is in the repulsive region).
The neutral particles called  breather emerge.
The mass spectra of the breathers are known \cite{SG}:
  \beq
\label{eq:spectrum}
      m_{\ell}=2m\sin(\frac{\pi}{2}\gamma \ell)
	     \; \; \, \;\;\; \ell=1,2,\cdots< \frac{1}{\gamma}\equiv
\frac{8\pi}{\beta^{2}}-1.
  \eeq
 Here $m$ is  the mass of the soliton.
For our case $\frac{1}{\gamma}=n+\frac{1}{2}$,
$n$ species of  breathers with mass $m_{\ell}=2m\sin(  \frac{\pi \ell}
 {2n+1})$  emerge.
The location of the pole of the $\ell$-th breather is
  \beq
     i\alpha_{\ell}\pi=i\pi\left(1-\frac{2 \ell}{2n+1}\right).
  \eeq
Let
  \beq
 \label{eq:largeF}
    F_{\alpha}(v)  =
      \frac{\sinh v+i\sin\alpha\pi}
           {\sinh v-i\sin\alpha\pi}=
      \frac{\tanh\frac{1}{2}(v+i\alpha\pi)}
           {\tanh\frac{1}{2}(v-i\alpha\pi)}\;\;\;,
  \eeq
  which is crossing symmetric and
 has a simple pole at  $v=i\alpha\pi$ and at
$v=i(1-\alpha)\pi$.   The residues  are
$2i\tan\alpha\pi$ and $-2i\tan\alpha\pi$ respectively.
The  complete $Z$ factor is  thus
  \beq
     Z(v)=Z_{min}(v)\prod_{\ell=1}^{n}F_{\alpha_{\ell}}(v)\;\;.
  \eeq
Using the expression
  \beq
     F_{\alpha}(v)=
      \frac{\Gamma(\frac{1+\alpha}{2}+\frac{v}{2\pi i})
              \Gamma(\frac{1-\alpha}{2}-\frac{v}{2\pi i})
              \Gamma(\frac{\alpha}{2}-\frac{v}{2\pi i})
              \Gamma(1-\frac{\alpha}{2}+\frac{v}{2\pi i})  }
           {\Gamma(\frac{1+\alpha}{2}-\frac{v}{2\pi i})
              \Gamma(\frac{1-\alpha}{2}+\frac{v}{2\pi i})
              \Gamma(\frac{\alpha}{2}+\frac{v}{2\pi i})
              \Gamma(1-\frac{\alpha}{2}-\frac{v}{2\pi i})  }
  \eeq
and Malmsten's formula and changing the integration variable
  from $t$ to $(2n+1)t$,
we obtain the following result
  \beq
     Z(v)=\frac{(-1)^{n}}{\cosh(n+\frac{1}{2})v}
        \exp\left(  -i\int_{0}^{\infty}\frac{dt}{t}
               \frac{\sin((n+\frac{1}{2})v t/\pi)\sinh
 \left((n-\frac{1}{2})t/2 \right) }
                    {\sinh t/2\cosh \left( (n+\frac{1}{2})t/2 \right) }
            \right).
  \eeq
 The choice of the sign  factor  conforms
  to the case of
 the ordinary Sine-Gordon model  at
$\frac{\beta^{2}}{8\pi}=\frac{1}{n+3/2}$.

As we said before,
 the S-matrices for the breathers  and those
 for the breathers-(anti-)soliton are the same  as that of
  the ordinary Sine-Gordon.
The soliton-breather S-matrices $S_{as}(v)
 = S_{a \bar{s}}(v)$ are  \cite{KaTh}
  \beq
     S_{as}(v)=\prod_{r=0}^{a-1}F_{\frac{1-\gamma(a-2r)}{2}}(v)\;\;\;.
  \eeq
The breather-breather S-matrices are \cite{KaTh}
  \beq
\label{eq:bb}
     S_{aa'}(v)=F_{\frac{\gamma}{2}(a+a')}(v)
                       \prod_{r=1}^{a-1}F_{\frac{\gamma}{2}(a+a'-2r)}(v)

\prod_{r=1}^{a'-1}F_{\frac{\gamma}{2}(a+a'-2r)}(v)\;\;\;.
  \eeq

\section{  TBA with Reflections and the Diagonalization }

In the last section, we have constructed  the $S$ matrices which
 consist of  the soliton-antisoliton sector,
 the soliton-breather sector and the breather-breather sector.   The
  soliton-(anti-)soliton sector has the $S$ matrices which are nondiagonal
 while the second and the third sectors have only diagonal scattering.
The thermodynamic Bethe ansatz (TBA)  of ref. \cite{ALZ}
 is a method of computing  thermodynamic quantities of the system
  from its  physical S-matrices.  The case in which   S-matrices are diagonal
 has been discussed repeatedly and  we  just
 summarize the minimum content in Appendix $B$.
When the scattering is not diagonal,  the diagonalization of the product
 of the $S$ matrices is required, which is mathematically equivalent to the
 diagonalization of the corresponding transfer matrix.
The algebraic Bethe ansatz
  or  the quantum inverse
scattering method \cite{QIM}
 is a method of diagonalizing  a transfer
  matrix
without referring explicitly to the wave function of the system.
 We apply this  to the product of the $S$ matrices  constructed in the last
section.
 In what follows, we generically denote by
 \beq
     P_{\kappa}(v)
              = \lim_{L \rightarrow \infty}
 \frac{n_{i+1}^{(\kappa)}-n_{i}^{(\kappa)}}{L(v_{i+1}-v_{i})},
  \eeq
   the density of states of spieces $\kappa$  per unit rapidity and unit
  length and by
  \beq
     \rho_{\kappa}(v) =  \lim_{L \rightarrow \infty}
  \frac{1}{L(v_{i+1}-v_{i})}
  \eeq
   the density of  particles of spieces $\kappa$.  Here, $L$ is the size of
 the box and the number of particles in the system
    is denoted by  ${\cal N}$.
   The integer $n_{i}^{(\kappa)}$  labels the $i$-th level belonging to
 the species $\kappa$.      We introduce $\epsilon_{\kappa}(v) -\frac
  {\mu_{\kappa}}{T} $ by
  \beq
     \frac{\rho_{\kappa}(v )}{P_{\kappa}(v )} \equiv
     \frac{e^{ \frac{ \mu_{\kappa}}{T}-\epsilon_{\kappa}(v )}}
          {1+e^{  \frac{ \mu_{\kappa}}{T} -\epsilon_{\kappa}(v )}}.
 \label{eq:defepsilon}
  \eeq

  Let us first  diagonalize  the quantity
 \beq
 \label{eq:prod}
 (T_{ab}(v))_{c_{i}}^{d_{i}}\equiv \sum_{k_{i}}
        S_{ac_{1}}^{d_{1}k_{1}}(v-v_{1})
        S_{k_{1}c_{2}}^{d_{2}k_{2}}(v-v_{2}) \cdots
        S_{k_{{\cal N}-1}c_{ {\cal N} }}^{d_{{\cal N} }~~~~b}(v-v_{{\cal N}
   })\;\;\;
  \eeq
  associated with  the periodic boundary condition for the system
  \beq
 \label{eq:pbc1}
     e^{imL\sinh v_{i}} T_{ab}(v_{i})\Psi_{b}=\Psi_{a} \;\;\;
  \eeq
  in the infinite volume limit $ {\cal N}, \;L  \rightarrow \infty$.
{}From the unitarity, we see that
 $S_{ab}^{cd}(0)=\pm \delta_{a}^{d}\delta_{b}^{c}$
  holds.
Eq.~(\ref{eq:pbc1}) can then be written as
  \beq
     e^{im L\sinh v_{k}} [ \pm \sum_{b}T_{bb}(v=v_{k})]
  \Psi_{a} = \Psi_{a} \;\;\;.
  \eeq
  The quantity ${\displaystyle \sum_{b}T_{bb}(v)}$
 is called transfer matrix.
 The $S$ matrix seen in eq.~(\ref{eq:prod}) is taken to be
   our S-matrix   of eq.~(\ref{eq:Smatrix})
 describing the scattering of  $u$, $d$  solitons
  \beq
     \bordermatrix{    & du & ud \cr
                    ud & b  & \tilde{c} \cr
                    du & c  & \tilde{b} \cr
                                             },\ \ \
     \bordermatrix{    & uu & dd \cr
                    uu & a  & 0         \cr
                    dd & 0  & \tilde{a} \cr  } \;\;\;,
  \eeq
  with $a= \tilde{a}$, $b=\tilde{b}$ and $c= \tilde{c}$.
  For notational simplicity, we have  identified  $ \bar{d} =u$ and
  $ \bar{u} = d$.   The diagonalization in this section   does not lose
 generality by this identification.
This S-matrix satisfies the condition
  \beq
     a(v)\tilde{a}(v)+b(v)\tilde{b}(v)-
     c(v)\tilde{c}(v)=0\;\;\;, \label{eq:ffer}
  \eeq
  which is nothing but a free fermion condition for the six vertex  model.
The  operator defined by
  \beq
  \label{eq:mono}
     T_{ab}(v)=\bordermatrix{      & u         & d         \cr
                                      u & A(v) & B(v) \cr
                                      d & C(v) & D(v) \cr   }
  \eeq
 is called monodromy operator.
Here $A, \ B, \ C$ and $D$ are $2^{{\cal N}}\times 2^{ {\cal N} }$ matrices.
The reference state $ \mid \Omega >$ is  a state in which  all of
 the ${\cal N}$ particles
are  $d$-particles:
\beqn
  \mid  \Omega >~ \equiv~ \mid d(v_{1})d(v_{2})\cdots d(v_{ {\cal N} })>
\;\;\;.
\eeqn
  The state  $ \mid \Omega >$
   satisfies $C(v) \mid \Omega > = 0$ for any $v$,
  and  is  found to be  an eigenstate
 of  both $A(v)$ and $D(v)$,
  \beq
     A(v) \mid \Omega >~ =~ \prod_{i=1}^{ {\cal N} }b(v-v_{i}) \mid
 \Omega >\;\;,
\ \ \
     D(v) \mid \Omega >~ =~ \prod_{i=1}^{  {\cal N} }
  \tilde{a}(v-v_{i}) \mid \Omega
  > \;\;\;,
  \eeq
  and, therefore, that of the transfer matrix  $A(v)+D(v)$.

It is known that
the eigenstates of the transfer matrix  can be
   constructed by  successively applying  the operator $B(v)$ on
  $ \mid \Omega >$:
  \beq
     \psi_{n^{\prime}} \left( \{\{ y_{r} \}\} \right)
    =\prod_{r=1}^{n^{\prime}}B(y_{r}) \mid \Omega  > \;\;\;
  \eeq
 Here   the set of rapidities  $ \{ \{ y_{r}\} \}$ is
 subject to the periodic boundary condition.
The above state contains $ n^{\prime}$ $u$-particles and
 $N-n^{\prime}$ $d$-particles.
{}From the Yang-Baxter relation for the S-matrices
  \beq
     S_{ab}^{a'b'}(v-y_{r})S_{b'c}^{b''d}(v)S_{a'b''}^{fe}(y_{r})=
     S_{bc}^{b'c'}(y_{r})S_{ab'}^{fb''}(v)S_{b''c'}^{ed}(v-y_{r}),
  \eeq
one can show that the transfer matrix obeys the following relations
  \beq
     S_{ac}^{a'c'}(v-y_{r})T_{c'd}(v)T_{a'b}(y_{r})=
     T_{cd'}(y_{r})T_{ab'}(v)S_{b'd'}^{bd}(v-y_{r}).
  \eeq
{}From these relations for (abcd)=$(uuud),(uduu),(ddud)$ and
 eq.~(\ref{eq:ffer}), we find
  \beqn
     \left\{ \begin{array}{l}
         A(v)B(y_{r})=
              -\frac{\tilde{a}(v-y_{r})}{\tilde{b}(v-y_{r})}
				    B(y_{r})A(v)
              +\frac{c(v-y_{r})}{\tilde{b}(v-y_{r})}
				    B(v)A(y_{r}), \\
         D(v)B(y_{r})=
              +\frac{\tilde{a}(v-y_{r})}{\tilde{b}(v-y_{r})}
				                B(y_{r})D(v)
              -\frac{c(v-y_{r})}{\tilde{b}(v-y_{r})}
				    B(v)D(y_{r}).
     \end{array} \right.
  \eeqn
   as well as
  \beqn
     [A(v)+D(v)]\psi &=& \prod_{r=1}^{n^{\prime}}
       \frac{\tilde{a}(v-y_{r})}{\tilde{b}(v-y_{r})}
       \left((-1)^{n^{\prime}}\prod_{i=1}^{  {\cal N} }b(v-v_{i})
       +\prod_{i=1}^{  {\cal N} }\tilde{a}(v-v_{i})\right)\psi   \\
    &+& ( {\rm  unwanted~terms})\;\;\;.  \nonumber
  \eeqn
The unwanted terms are  found to be
 proportional to $[(-1)^{n^{\prime}}A(y_{r})+D(y_{r})] \mid \Omega > $ and
 vanish if
  \beq
     \prod_{i=1}^{ {\cal N} }\frac{b(y_{r}-v_{i})}{\tilde{a}(y_{r}-v_{i})}=
          (-1)^{n^{\prime}+1}  \;\;\;. \label{eq:yr}
  \eeq
Eq.~(\ref{eq:yr}) is nothing but a periodic boundary condition  which
 determines  a set of allowed  rapidities $y_{r}$.
 The eigenvalues of the transfer matrix are then given by
  \beq
     \lambda^{(n^{\prime})} (v; \{\{ y_{r} \}\})=\prod_{r=1}^{n^{\prime}}
      \frac{\tilde{a}(v-y_{r})}{\tilde{b}(v-y_{r})}
       \left(
	         (-1)^{n^{\prime} }\prod_{i=1}^{  {\cal N} }b(v-v_{i})
             +\prod_{i=1}^{ {\cal N} }\tilde{a}(v-v_{i})
	   \right). \label{eq:eigenlam}
  \eeq
where $y_{r}$ are the roots of  eq.~(\ref{eq:yr}).

To summarize,
the eigenvalues of the transfer matrix are
  \beqn
 \label{eq:ev}
     \lambda^{(n^{\prime})} (v;\{\{ y_{r}\}\} )
   &=& C\prod_{i=1}^{{\cal N}}Z(v-v_{i})
     \prod_{r=1}^{n^{\prime}}  (-1)^{(n-\ell)}i \cosh \left( (n
+\frac{1}{2})(v-y_{r}) \right)   \nonumber \\
   & \times&  \prod_{r=n^{\prime}+1}^{  {\cal N} }
\sinh \left((n+\frac{1}{2})(v-y_{r}) \right)
  , \label{eq:lamd2}
  \eeqn
where $\{\{ y_{r}\}\} $ are the roots of the following equations
  \beq
      \prod_{i=1}^{ {\cal N} }\frac{i(-1)^{(n-\ell+1)}
 \sinh(n+\frac{1}{2})(y
_{r}-v_{i})}
      {\cosh(n+\frac{1}{2})(y_{r}-v_{i})}=(-1)^{n^{\prime}+1} \;\;\;.
 \label{eq:yreq2}
  \eeq
Here
  \beqn
     C \equiv \prod_{r=1}^{ {\cal N} } \left( \sinh (n+\frac{1}{2})(v-y_{r})
   \right)^{-1}
     \left[
       \prod_{i=1}^{  {\cal N} }\left( \cosh(n+\frac{1}{2})(v-v_{i}) \right)
    \right. \nonumber \\
   \left. +(-1)^{n^{\prime} }
       \prod_{i=1}^{{\cal N}} (-i)\sinh  \left((n+\frac{1}{2})(v-v_{i}) \right)
     \right]\;\;.
  \eeqn
Note that, for $\{\{y_{r}\}\}$ satisfying  eq.~(\ref{eq:yreq2}),
 $C$ has no pole in $v$
and is bounded at infinity.  It must therefore be independent of $v$.

In the physical strip, the roots of eq.~(\ref{eq:yreq2}) are
  \beq
  \label{eq:modes}
     \left\{ \begin{array}{l}
         y_{r}^{(k)}=z_{r} + \frac{k+\frac{1}{2}}{2n+1}\pi i\ \ \ \ \
           k=0,1,2,\cdots  2n\;\;, \\
         y_{r}^{(\bar{k})}=z_{r} -
           \frac{\bar{k}+\frac{1}{2}}{2n+1}\pi i\ \ \ \ \
           \bar{k}=0,1,2,\cdots  2n\;\;.
     \end{array} \right.    \;\;\;
  \eeq
Here $z_{r}^{,} s$ are real numbers.  Let us denote
  these $4n+2$  modes  in eq.~(\ref{eq:modes})   by
\beqn
\label{eq:modelabel}
 \{\{ {\bf 0,1,2,\cdots,2n  \}\}  \oplus \{\{  \bar{0},\bar{1},\bar{2},\cdots,
  \overline{2n}}   \}\} \equiv {\cal L}  {\bf \oplus}
 \overline{\cal L}\;\;\;.
\eeqn

The density of states for these $4n+2$ modes is denoted by
$P^{(\ell)}(z^{(\ell)})~~  {\rm for}\;\;\;\ell \in   {\cal L} {\bf \oplus}
\overline{\cal L}.$
 From eq.~(\ref{eq:yreq2}), we  find
  \beq
\label{eq:ssformula}
      P^{(\ell)}(z^{(\ell)})=(-1)^{\ell}(2n+1)\int \frac{dv}{2\pi}
  \frac{\rho_{\bf m}(v)}
      {\cosh(2n+1)(z^{(\ell)}-v)}\
 \;\;\; {\rm for} \; \ell  \in   {\cal L}  {\bf \oplus} \overline{\cal L}
 \;\;\;, \label{eq:mlpbc}
  \eeq
  in the $ {\cal N} \rightarrow \infty, L \rightarrow \infty$ limit.
  Here,  $\rho_{\bf m}(z)  \equiv
  {\displaystyle \lim_{ L \rightarrow
 \infty }  \frac{1}{ L (  y_{r+1}^{(k)}
 - y_{r}^{(k)} ) }   } $
 has no dependence on the superscript $(k)$.
  We refer to this as the density of particles ( occupancy)
 for the massive mode ${\bf m}$.  On the other hand,
comparing eq.~(\ref{eq:ssformula})
 with the periodic boundary condition eq.~(\ref{eq:Rho}),
  we see that  the
  $4n+2$ modes  lying along the purely imaginary direction
 are interpreted  as massless modes appearing from the soliton
 distributions.   The semi-classical treatment of these modes
  should  correspond to a generalization of the discussion
  of the fractional fermion number
  in \cite{JR,SSH}.

  We now turn to
the  (anti-)soliton-breather S-matrices $S_{as} \left( = S_{a \bar{s}}
 \right)$
 which are diagonal.
Using the  monodromy operator $T(v)$  of eq.~(\ref{eq:mono}),
 the periodic boundary condition
for the soliton with mass $m$ reads
  \beq
     e^{imL\sinh v}\prod_{j}S_{a s}(v-v_{j})T(v)\Psi=\Psi \;\;\;.
  \eeq
{}From this equation,
  \beq
  \label{eq:pbcas}
    \frac{m}{2\pi} \cosh v + \lim_{L \rightarrow \infty}
  \frac{1}{L}{\Im}\frac{d}{dv}\log\lambda^{(n^{\prime})} (v; \{\{ y_{r} \}\}  )
          +\sum_{a=1}^{n} \int  \frac{dv^{\prime}}{2 \pi}
 \phi_{a {\bf m} }(v-v^{\prime})\rho_{a}(v^{\prime})
	 =  P_{\bf m}(v)  \;\;\;,
  \eeq
where $\phi_{a  {\bf m}}(v) = -i(\log S_{as})'$, $\rho_{a} (v)$
 is the density of particles for
the $a$-th breather, and $P_{\bf m}(v)$ is
 the density of states for ${\bf m}$.

We would like to express ${\displaystyle  \lim_{L \rightarrow \infty}
  \frac{1}{L}{\Im}\frac{d}{dv}\log\lambda(v)^{(n^{\prime})} (v; \{\{ y_{r} \}\}
 )
    }$ in terms of the densities
 of  the $(4n+2)$  massless  modes  in  eq.~(\ref{eq:modelabel}).
  For that purpose, we note that
the $2^{{\cal N} }$ eigenvalues of
 the transfer matrix $\lambda^{(n^{\prime})}(v,\{\{ y_{r} \}\} )$
$n^{\prime} = 1, \cdots {\cal N}$    in eq.~(\ref{eq:ev}) are
determined by   choosing
 between $\sinh \left( (n+\frac{1}{2})(v-y_{r}) \right)$ and
$\cosh \left( (n+\frac{1}{2})(v-y_{r}) \right)$   for each
$y_{r}$.
 We define the density
$\rho_{-}^{(\ell)}(z)\;\;$  $ \ell = k (~ {\rm or}~
 \bar{k})
 \in {\cal L}  ( ~ {\rm or}~ \overline{\cal L} )$ by the occupancy
 of the mode $k (~ {\rm or}~  \bar{k})$
 at $y^{(k)} = z + \frac{k+ \frac{1}{2} }{2n+1} \pi i$
    (or    $y^{({\bar k})} = z - \frac{{\bar k}+ \frac{1}{2} }{2n+1} \pi i$ )
  which contributes a $\cosh$ to
   $\lambda^{(n^{\prime})}(v,\{\{ y_{r} \}\} )$ for
  $\ell =k$ (or a $\sinh$
for $\ell=\bar{k}$).  Similarly, we define
the density $\rho_{+}^{(\ell)}(z )$  by the occupancy of the
 mode $k =\ell (~ {\rm or }~ \bar{\ell} )$
 which contributes a $\sinh$
   ( or a $\cosh$.)
By construction,
\beq
 P^{(\ell)} (v) = \rho_{+}^{(\ell)} (v) + \rho_{-}^{(\ell)} (v)\;\;\;,
\eeq
   which permits us to think of $+$ modes as ``particles'' and $-$ modes as
  ``holes''.   From eq.~(\ref{eq:ev}),
 we  find, after some calculation,
  \beqn
  \label{eq:loglambda}
      \lim_{L \rightarrow \infty}
   \frac{1}{L}{\Im}\frac{d}{dv}\log\lambda(v)  &=&
     \int dv^{\prime}\rho(v^{\prime})
{\Im}\frac{d}{dv}\log Z(v-v^{\prime})  \nonumber \\
       &+&  \left( n+\frac{1}{2} \right)
 \sum_{\ell =k,\bar{k}}\int dv^{\prime}
     \frac{(-1)^{\ell}(\rho_{+}^{(\ell)}( v^{\prime})-
  \rho_{-}^{(\ell)}(v^{\prime})     )}
          {\cosh(2n+1)(v-v^{\prime})} \;\;.
  \eeqn
    We need in turn to  compute  ${\Im}\frac{d}{dv}\log Z(v-v^{\prime})$.
{}From our  S-matrices  in eq.~(\ref{eq:Smatrix2}), we  find
  \beq
     {\Im}\frac{d}{dv}
     \log Z(v)=-(n+\frac{1}{2})\int dt e^{i(2n+1)v t}
	   \frac{\sinh(n-\frac{1}{2})\pi t}
            {2\sinh\frac{\pi}{2}t\cosh\frac{\pi}{2}t\cosh(n+\frac{1}{2})\pi
t}\;\;\;.
  \eeq
Using the formulas concerning with the Fourier transform
  \beqn
     \frac{1}{\sinh pt}  &=& \frac{i}{2p}\int
 dx e^{-ixt}\tanh \frac{\pi x}{2p}\;\;\;, \\
     \frac{1}{\cosh pt}  &=&
	 \frac{1}{2p}\int dy e^{-iyt}\frac{1}{\cosh \frac{\pi y}{2p}}\;\;\;, \\
    {\rm and}~~~~~  \frac{\sinh pt}{\cosh \pi t}  &=&
\frac{i}{\pi}\int dz e^{-izt}
     \frac{\sin \frac{p}{2}\sinh\frac{y}{2}}{\cosh z +\cos p}\;\;\;,
  \eeqn
we obtain
  \beqn
   &~&  {\Im}\frac{d}{dv}\log Z(v)  \nonumber \\
  &=&(n+\frac{1}{2})\int\frac{dx}{\pi}
     \int\frac{dz}{\pi}
	 \frac{\tanh x\sin  \left(\frac{n-\frac{1}{2}}{2n+1}\pi\right) \sinh
\frac{z}{2}}
          {\cosh  \left( (2n+1)(v-\frac{z}{2}-\frac{x}{2n+1}) \right)
                                  (\cosh z+\cos \frac{2n-1}{2n+1}\pi)} \;\;\;.
  \eeqn
  After a change of  variables from $z$ to $z'=\frac{z}{2}+\frac{x}{2n+1}$,
this equals
  \beqn
     (2n+1)\int \frac{dz'}{\pi}\frac{1}{\cosh(2n+1)(v-z')}
     \int \frac{dx}{\pi}
	 \frac{ \tanh x  \sin  \left(\frac{n-\frac{1}{2}}{2n+1}\pi \right) \sinh
(z'-\frac{x}{2n+1})}
          {\cosh(2z'-\frac{2x}{2n+1})+\cos \left( \frac{2n-1}{2n+1}\pi  \right)
 } \;\;\;.
  \eeqn
   Carrying out  inverse Fourier transforms
  \beqn
     \tanh x &=& -\frac{i}{2}\int
 dy e^{ixy}\frac{1}{\sinh\frac{\pi}{2}y}\;\;\;,
    \\
     \frac{\sin\frac{p}{2}\sinh\frac{y}{2}}{\cosh y+\cos p}  &=&
	 -\frac{i}{2}\int dw e^{iyw}\frac{\sinh pw}{\cosh \pi w}\;\;\;,
  \eeqn
we obtain the final form for $ {\Im}\frac{d}{dv}\log Z(v)$:
  \begin{eqnarray}
 \label{eq:logz}
   &~&  {\Im}\frac{d}{dv}\log Z(v)  \nonumber \\
  &=&     -(2n+1)\int\frac{dz}{2\pi}\frac{1}{\cosh(2n+1)(v-z)}
     \int dw e^{2izw}\frac{\sinh\frac{2n-1}{2n+1}\pi w}
                          {\sinh\frac{\pi}{2n+1}w\cosh\pi w} \nonumber \\
    &=&-(n+\frac{1}{2})\int
      dz\frac{1}{\cosh(2n+1)(v-z)}\sum_{m=-n+1}^{n-1}
      \frac{1}{\cosh(z+\frac{m\pi i}{2n+1})}\;.
  \end{eqnarray}

  Putting eqs.~(\ref{eq:pbcas}),~(\ref{eq:loglambda}) and (\ref{eq:logz})  all
together,
 we  finally obtain
  \begin{eqnarray}
 \label{eq:bseq}
       \frac{m}{2\pi}    \cosh v&+&\sum_{ \ell  \in {\cal L}
 {\bf \oplus} \overline{\cal L} }
 \int  \frac{dv^{\prime}}{2\pi} \left\{
     \rho_{+}^{(\ell)}(v^{\prime})\phi_{+}^{(\ell)}(v- v^{\prime}   )+
     \rho_{-}^{(\ell)}(   v^{\prime})
 \phi_{-}^{(\ell)}(v-  v^{\prime})\right\} \nonumber \\
     &+&\sum_{a=1}^{n}\int
  \frac{dv^{\prime}}{2\pi}  \phi_{a {\bf m}}(v- v^{\prime})
  \rho_{a}(v^{\prime})=
      P_{{\bf m} }(v)  \;\;\;,               \label{eq:mmpbc}
  \end{eqnarray}
    which have originated from the periodic boundary condition
 for the massive mode.
 We have denoted by $\phi_{\pm}^{(\ell)}$
  \beq
     \phi_{\pm}^{(\ell)}(v)\equiv -\frac{(-1)^{\ell}}{4n+2}
     \sum_{m}\frac{1}{\cosh(v+\frac{m\pi i}{2n+1})}
     \pm(-1)^{\ell} \left( n+\frac{1}{2} \right)
 \frac{1}{\cosh(2n+1)v}  \;\;\;.
  \eeq

  So far, we have obtained  a system of nonlinear equations from
 soliton-antisoliton scattering
  (eq.~(\ref{eq:ssformula})) and from
 breather-soliton scattering (eq.~(\ref{eq:bseq})).  As for the
breather-breather scattering which is diagonal,
  we have eqs.~(\ref{eq:dia}) and (\ref{eq:phi}) with the $S$-matrices taken
from
 eq.~(\ref{eq:bb}).
In order to derive a closed set of  nonlinear equations,  we have to minimize a
free energy with chemical potentials under the conditions
eqs.~(\ref{eq:ssformula}), (\ref{eq:bseq}) and (\ref{eq:dia}).  Having done
this, we find
  a system of nonlinear  equations  which
 determine   $\epsilon_{\lambda}(v)$    as functions
  of temperature and chemical potentials:
  \beqn
\label{eq:epseq}
     \epsilon_{\lambda}(v )  &=&  \frac{ m_{\lambda}}{T}  \cosh v -
     \sum_{
  \kappa = {\bf m}, \ell, a } \int
	 \frac{dv^{\prime}}{2\pi}\phi_{\kappa\lambda}(v^{\prime}-v)
     \log(1+e^{  \frac{ \mu_{\kappa}}{T} - \epsilon_{\kappa}(v^{\prime})})
 \;\;\;,   \\
  {\rm where }~~~&~&   \nonumber \\
     \phi_{ab}(v)   &=&  -4\cosh v
     \left( \sum_{r=0}^{a-1}+\sum_{r=1}^{b-1}\right)
     \frac{\sin\frac{(a+b-2r)\pi}{2n+1}}
	      {\cosh2v-\cos\frac{(a+b-2r)2\pi}{2n+1}} \;\;\;,  \\
     \phi_{ {\bf m}a}(v)  &=&   \phi_{a {\bf m} }(v)=-4\cosh v\sum_{r=0}^{a-1}
     \frac{\sin(\frac{1}{2}-\frac{a+2r}{2n+1})\pi}
	      {\cosh 2v-\cos(\frac{1}{2}-\frac{a+2r}{2n+1})2\pi} \;\;\;,  \\
     \phi_{\ell {\bf m}}(v)   &=&   \phi_{ {\bf m} \ell}(v)=
	 (2n+1)\frac{(-1)^{\ell}}{\cosh(2n+1)v} \;\;\;, \\
     \phi_{  {\bf m m} }(v)   &=&   \sum_{ \ell  \in {\cal L} \oplus
  \overline{\cal L} }\int \frac{dv'}{2\pi}
     \phi_{-}^{(\ell)}(v-v')
     \phi_{\ell  {\bf m}  }(v')   \;\;\;.
  \eeqn

\section{ The  Massless Limit}

   We now proceed to  take the massless limit
 of the system  and compute the
central charge $\tilde{c}$ in  the  limit.
When $m\rightarrow 0$, the only nonvanishing contributions for
 eq.~  (\ref{eq:epseq}) ( or~eq.~(\ref{eq:TBAeq}))
 come from the region of integration at $\mid v \mid
\rightarrow \infty$.
We may therefore replace $\epsilon_{a}(v)$ and $\rho_{a}(v)$ by their
 asymptotic forms.
{}From  eq.~(\ref{eq:Rho}) we have
  \beq
     P_{a}(v) \approx \frac{m_{a}}{4\pi}e^{\mid v \mid}
  \;\;,\ \ \
     \mid v \mid \rightarrow\infty  \;\;\;,
  \eeq
and, from  eq.~(\ref{eq:defep}), we find
  \beq
     \rho_{a}(v)\approx \frac{m_{a}}{4\pi}
     \frac{e^{\mid v \mid}}
	      {1+e^{\epsilon_{a}(v) - \frac{\mu_{a}}{T}
  }}\;\;,\ \ \
     \mid v \mid \rightarrow\infty \;\;\;.
  \eeq
Taking the derivative of  eq.~(\ref{eq:epseq}) or
 (\ref{eq:TBAeq}) gives, at large $v$,
  \beq
     \frac{d\epsilon_{a}(v)}{dv}\approx  \frac{ m_{a}}{T}\sinh v
     \approx (sgn\ v)\frac{1}{2}  \frac{m_{a}}{T} e^{\mid v \mid}\;\;\;,\ \ \
     \mid v \mid \rightarrow\infty \;\;\;.
  \eeq
Then the entropy  (eq.~(\ref{eq:ent})) becomes
  \begin{eqnarray}
     \lim_{m\rightarrow 0}S
	 &=&\frac{T}{2\pi}\sum_{a}\int_{-\infty}^{\infty}
        dv\frac{d\epsilon_{a}}{dv}(sgn v)
        \left[ \log(1+e^{\epsilon_{a}-\frac{\mu_{a}}{T} })-
           \frac{\epsilon_{a}-\frac{\mu_{a}}{T} }
{1+e^{\frac{\mu_{a}}{T}-\epsilon_{a}}}
		\right] \nonumber \\
     &=&\frac{T}{2\pi}\sum_{a}
        \left\{ \int_{\epsilon_{a}(0)}^{\epsilon_{a}(\infty)}-
                \int_{\epsilon_{a}(-\infty)}^{\epsilon_{a}(0)}
		\right\} d\epsilon_{a}
        \left[  \log(1+e^{\epsilon_{a}-\frac{\mu_{a}}{T} })-
            \frac{\epsilon_{a}-\frac{\mu_{a}}{T}}
{1+e^{\frac{\mu_{a}}{T}-\epsilon_{a}}}
	    \right].
  \end{eqnarray}
Using the evenness of the functions $\epsilon_{\kappa}(v)$,
and making a change of variables
  \beq
     f(\epsilon_{\kappa}-  \frac{\mu_{\kappa}}{T}   )
	 \equiv \frac{1}{1+e^{\epsilon_{\kappa}-
  \frac{ \mu_{\kappa}}{T}  }}  \;\;\;,
  \eeq
   we find
  \beq
     \lim_{m/T \rightarrow 0}S=\frac{T}{\pi}\sum_{\kappa}
     \int_{f(\epsilon_{\kappa}(0)- \mu_{\kappa}/T )}
	     ^{f(\epsilon_{\kappa}(\infty)- \mu_{\kappa}/T )}
        df\left[ \frac{\log f}{1-f}+\frac{\log(1-f)}{f}\right] \;\;\;.
  \eeq
  (See, for example,  the second entry of \cite{IM}  as well as
 \cite{BR,KR}.)
This integral is essentially the Rogers dilogarithm. We have collected
    useful formulas associated with dilogarithm in
 \ref{sec:Rdil} C.

Let the chemical potential for $\kappa$ be
 $\mu_{\kappa}=i \alpha_{\kappa}T $
 and let the entropy per unit length be
  \beq
     S=\frac{\pi\tilde{c}(\frac{m}{T})}{3}T    \;\;\;.
  \eeq
  Then
\beqn
 \label{eq:cuv}
 \tilde{c} &\equiv&
 \lim_{m/T\rightarrow 0}\tilde{c}(\frac{m}{T}) \;\;\;, \nonumber  \\
      &=&   \frac{6}{\pi^{2}}\sum_{\kappa}
     \left( L[f(\epsilon_{\kappa}(0)- i \alpha_{\kappa})]-
     L[f(\epsilon_{\kappa}(\infty)-
 i \alpha_{\kappa})]\right)
\equiv   \tilde{c}_{0} -  \tilde{c}_{\infty}  \;\;\;.
  \eeqn
 Ref.\cite{FSC} tells us $\tilde{c} = c-12 \left( h_{{\rm min}}
 +\bar{h}_{ {\rm min} }  \right)$ with
$c$ the central charge of the corresponding conformal field theory in the
massless limit and $h_{ {\rm min}}+ \bar{h}_{ {\rm min}}$
 the lowest scaling dimension.
 The function $L(x)$ is defined in eq.~(\ref{eq:dilog}).

  How to determine $\epsilon_{\kappa}(0)$ and
$\epsilon_{\kappa}(\infty)$ is well-known by now. A general feature of
the solution $\epsilon_{\kappa}(v)$ is that as $r= \frac{m}{T}
\rightarrow 0$ it has a plateau of length $\sim \log(1/r)$ around the
origin.  Taking the limit $r\rightarrow 0$ and $v\rightarrow 0$, the
first term in the right hand side of eq.~(\ref{eq:epseq}) or
(\ref{eq:TBAeq}) can be neglected.  Thus $\epsilon_{\kappa}(0)$
satisfies the algebraic equations \beq
\label{eq:eps0}
     e^{\epsilon_{\kappa}(0)}= \prod_{\kappa^{\prime} }\left(1+e^
{ \frac{\mu_{\kappa^{\prime}} }{T} -\epsilon_{\kappa^{\prime}
}(0)}\right)^{N_{\kappa \kappa^{\prime}}} \;\;\;, \eeq where \beq
N_{ab}=-\frac{1}{2\pi}\int_{-\infty} ^{\infty}dv \phi_{ab}(v)\;\;\;.
\eeq

  As for the other limit $r\rightarrow 0$, $v\rightarrow \infty$, only
the massless modes survive and \beqn e^{\epsilon_{\kappa}(\infty)} &=&
+\infty\;\;\;, \;\;\; \kappa = a, {\bf m}\;\;\;, \\
\label{eq:epsinfty}
     e^{\epsilon_{\kappa}(\infty)} &=&
\prod_{\kappa^{\prime}}\left(1+e^{ \frac{\mu_{\kappa^{\prime} } }{T}
-\epsilon_{ \kappa^{\prime}} (\infty) } \right )^{N_{\kappa
\kappa^{\prime} } }\;\;\;, \;\;\; \kappa = \ell \in {\cal L} \oplus
\overline{\cal L} \;\;\;.  \eeqn

We now turn to the evaluation of $N_{\kappa \kappa^{\prime}}$.  From
eq.~(\ref{eq:largeF}), we obtain \beq -\frac{1}{2\pi}\int dv(-i)(\log
F_{\alpha}(v))'=sgn\alpha\;\;\;.  \eeq We find \begin{eqnarray}
N_{ab}&=&\left\{ \begin{array}{l} a+b-1\ \ \ \ \ \ \ \mid
a-b\mid~=0,1, \\ 2 \min (a,b)\ \ \ \ \ \mid a-b\mid~ \geq 2\;\;.
\end{array} \right.  \nonumber \\ N_{a {\bf m} } &=& N_{{\bf m} a} = a
\;\;\;, \eeqn for the elements in which the breathers are involved.
The remaining elements are evaluated to be \beqn N_{\ell +} &=&
-\frac{1}{2}(-1)^{\ell} \frac{1}{2n+1}\;\;, \;\;\; N_{\ell -} =
\frac{1}{2}(-1)^{\ell} \frac{n}{2n+1}\;\;, \nonumber \\ N_{\ell {\bf
m} } &=&N_{ {\bf m} \ell} = -\frac{1}{2}(-1)^{\ell} \;\;\;, \nonumber
\\ N_{{\bf m} {\bf m} }&=&-\sum_{\ell \in {\cal L} \oplus
\overline{\cal L} }N_{\ell -} N_{\ell{\bf m}}	 =-\sum_{\ell \in
{\cal L} \oplus \overline{\cal L}
}\left((-1)^{\ell}\frac{n}{2n+1}\right)
\times\left(-\frac{1}{2}(-1)^{\ell}\right)=n\;\;\;.  \eeqn

 Let us solve eqs.~(\ref{eq:eps0}) and (\ref{eq:epsinfty}) for zero
chemical potentials.  Define $x_{\lambda}=e^{-\epsilon_{\lambda}(0)}$.
Eq.~(\ref{eq:eps0}) reads \begin{eqnarray}
\label{eq:alg}
     \frac{1}{x_{a}}&=&\prod_{b=1}^{a-1}(1+x_{b})^{2b}
(1+x_{a})^{2a-1}\prod_{c=a+1}^{n}(1+x_{c})^{2a} (1+x_{m})^{a} \;\;\;.
\nonumber \\ \frac{1}{x_{{\bf m}}}&=&\prod_{b=1}^{n}(1+x_{b})^{b}
(1+x_{ {\bf m} })^{n}(1+x_{even})^{-n-1} (1+x_{odd})^{n} \;\;\;.
\nonumber \\ \frac{1}{x_{even}} &=& x_{odd}= (1+x_{ {\bf m}
})^{-\frac{1}{2}}\;\;\;.  \end{eqnarray} Here we recall that the
indices $a, b$ refer to $n$ kinds of breathers while the subscript
${\bf m}$ denotes the massive mode.  The $4n+2$ massless modes of
${\cal L} \oplus \overline{\cal L}$ are divided into the $2n+2$ even
modes and the $2n$ odd modes.  After some guesswork for $n=1,2,3$
cases, we find a solution to eq.~(\ref{eq:alg}) : \begin{eqnarray}
e^{\epsilon_{a}(0)} = a(a+2) \;\;\;,\;\;\; e^{\epsilon_{ m}(0)} =
\frac{(n+1)^{2}}{2n+3}\;\;\;,\;\;\; e^{\epsilon_{even}(0)} =
e^{-\epsilon_{odd}(0)} = \frac{n+1}{n+2} \;\;\;, \end{eqnarray} which
yields
\beqn
\label{eq:c0}
  \tilde{c}_{0} = 2n+2 \;\;\;.
\eeqn

For $\epsilon_{\lambda}(\infty)$, the corresponding set of algebraic
equations is obtained by removing the massive mode ${\bf m}$ and the
breathers $a = 1 \sim n$ in eq.~(\ref{eq:alg}):
\beqn
\epsilon_{m}(\infty)=\infty \;\;\;,
 \;\;\; \epsilon_{a} (\infty) = \infty \;\;\;.
\eeqn
The solution is \begin{eqnarray}
e^{\epsilon_{even}(\infty)}=e^{\epsilon_{odd}(\infty)} = 1 \;\;\;,
\end{eqnarray} which means
\beqn
\label{eq:cinf}
 \tilde{c}_{\infty} = 2n+1 \;\;\;.
\eeqn
  In deriving eqs.~(\ref{eq:c0}) and (\ref{eq:cinf}), we have used
eqs.~(\ref{eq:L1}), (\ref{eq:Lm}).  We obtain, from
eq.~(\ref{eq:cuv}), \beq \tilde{c} = 1 \;\;,\; \;\;\; {\rm for~~ n=
0,1,2, \cdots} \;\;\;.  \eeq This is our main conclusion in this
section.

 We have also carried out calculation in the presence of purely
imaginary chemical potentials.  This is presented in Appendix D.  We
have determined the roots of the algebraic equation
eq.~(\ref{eq:eps0}) for the case of the fermion number $ f= 1/2$.
Having been unable to draw a physical conclusion from the calculation
unfortunately, we can state it only as a mathematical curiosity.

\section{ The  $c=1$ Conformal Field Theory}

  We now turn to identify the $c=1$ conformal field theory that our
model belongs to in the massless limit.  The known $c=1$ conformal
field theories ( see for example \cite{c=1rev} for review), except the
cases with isolated moduli \cite{Gin}, are characterized by the two
marginal directions, namely, the gaussian line and the orbifold line.
To fix our notation, we write a compactified free massless scalar and
the two point functions as $\varphi (z, \bar{z})= \frac{1}{2} \left
( \phi(z) + \bar{\phi}(\bar{z}) \right)$, $ < \phi(w) \phi(z)> = - \ln
(w-z)$, $ < \bar{\phi}(\bar{w}) \bar{\phi}(\bar{z})> = - \ln (\bar{w}-
\bar{z})$.  The radius parameter $r$ is introduced through $\varphi
(z, \bar{z}) \sim
\varphi (z, \bar{z}) + 2\pi r. $

 In the massless limit, the higher-spin $N=2$ supersymmetry algebra
derived in eqs.~(\ref{eq:SUSY}) and (\ref{eq:comm}) separates into the
unbar and the bar subalgebras: the central charges ${\cal Z}^{(\pm)}$,
which are spinless operators, become zero.  This can be done by
$m\rightarrow 0, v \rightarrow v + \Lambda,$ $ \Lambda \rightarrow +
\infty$, keeping $ me^{\Lambda}$ finite in eq.~(\ref{eq:action1})
  for $u(v)$ and $d(v)$ and
by $m\rightarrow 0, v \rightarrow v - \Lambda,$ $ \Lambda \rightarrow
+ \infty$, $ me^{\Lambda}$ finite in the corresponding formula for
 $\bar{u}(v)$ and $\bar{d}(v)$.  The
resulting algebra is regarded as a centerless subalgebra of a ${\bf
Z}_{2}$ graded chiral algebra for the $c=1$ conformal field theory.
In the case $n=0$, this chiral algebra is the well-known $N=2$
super-conformal algebra.  Focusing on the holomorphic part, we write
the limit as
\beqn
 Q^{\pm} \longrightarrow \int dz G^{\pm} (z) \;\;\;.
\eeqn
  Let us denote the conformal weights of the chiral operator
$G^{\pm}(z)$ by $(h_{G^{\pm}}~, \bar{h}_{G^{\pm}}=0)$.  This operator
can be represented by the chiral vertex operator
\beqn
\label{eq:cvo}
    G^{\pm}(z) = e^{\pm i p \phi(z)} \;\;\;, \;\;\; h_{G^{\pm}} =
\frac{p^{2}}{2} \;\;\;.
\eeqn
   Recall that the Lorentz spin of $ Q^{\pm}$ is $n+\frac{1}{2}$ at
off-criticality.  Taking the massless limit, we conclude that the spin
of $G^{\pm}(z)$ associated with the Eucliden rotation is $ n+
\frac{3}{2}$.  On the other hand, the primary fields of $U(1) \times
U(1)$ current algebra are a set of vertex operators $ V_{m
\ell}(z,\bar{z}) \equiv e^{ i p \phi(z) + i \bar{p}(\bar{z}) }$ with
the conformal weights
\beqn
\label{eq:prime}
  ( h, \bar{h}) &=& ( \frac{p^{2}}{2}, \frac{\bar{p}^{2}}{2} ) \;\;\;,
\;\;\; p = \frac{m}{2r} + \ell r\;\;\;, \;\;\; \bar{p} = \frac{m}{2r}
- \ell r\;\;\;, \;\;\ \\
\label{eq:ml}
  ( m, \ell ) &=& \{\{ m \in {\bf Z}, \ell \in {\bf Z}\}\} \;\;\;
\nonumber \\ {\rm or} ~~~~~ &=& \{\{ m \in {\bf 2Z}, \ell \in {\bf
Z}\}\} \oplus \{\{ m \in {\bf 2Z+1}, \ell \in {\bf Z+1/2} \}\} \;\;\;.
\eeqn
The spin is $m \ell$.  Eq.~(\ref{eq:ml}) is derived on the basis of
the mutual locality of $V_{m \ell}$ \cite{BPZ}.  In the latter case of
eq.~(\ref{eq:ml}) \cite{gauss2}, the GSO projection \cite{GSO} must be
given in order to render the spectrum modular invariant.

 Let us find the value of the radius parameter $r$.  For the chiral
vertex operator of eq.~(\ref{eq:cvo}), we should set $ \bar{h}=0$ in
eq.~(\ref{eq:prime}). This gives
\beqn
  r = \sqrt{ \frac{m}{2 \ell}} \;\;\;.
\eeqn
 From the spin, we find
\beqn
 h_{G^{\pm}} = m \ell = n+ \frac{3}{2} \;\;\;.
\eeqn
  We conclude that
\beqn
   r &=& \sqrt{2n +3} \;\;\;, \nonumber \\ m &=& \pm (2n + 3)
\;\;\;,\;\;\; \ell = \pm \frac{1}{2} \;\;\;. \;\;\;
\eeqn
   The model belongs to the latter possibiity of eq.~(\ref{eq:ml}).
The algebra generated by the operators eq.~(\ref{eq:cvo}) may be
called ${\bf Z}_{2}$ graded chiral algebra.  Closure of this algebra
tells us to include the chiral vertex operators $ e^{ iQ \phi(z)}$
with $Q$ integer multiple of $\sqrt{2n +3}$.  The operators
$v_{k}(z)=e^{i q_{k} \phi(z)}$ with $q_{k} = \frac{k}{\sqrt{2n+3}}, k=
0, 1, \cdots 2n +2$ provide representations of this algebra: $ [1],
[v_{1}],\cdots [v_{2n+2}]$.

    Now we turn to the question of which marginal line the model lies
in.  As in the case of the original Zamolodchikov $S$ matrices
\cite{ZZ}, the poles appearing in our soliton-antisoliton $S$ matrix
are interpreted as the ones in the direct channel of the
soliton-antisoliton scattering: they are understood as the bound state
poles of a soliton and an antisoliton.  The conservation of the $U(1)$
quantum number is thus seen.  We conclude that our model lies in the
gaussian line.

  A natural question arises. Can we have a freedom to change our $S$
matrices while maintaining the calculation and the discussion in the
previous sections, so that the model is converted to belong to the
orbifold line?  This is possible at the reflectionless points $
\frac{\beta^{2}}{8 \pi} = \frac{1}{n}\;,\;$ $ n= 2,3, \cdots$.  By
changing a sign of the $S$ matrix, we can make the bound state poles
appear in the crossed channel without changing the assumed mass
spectrum \cite{KM}.  The poles are then interpreted as the
soliton-soliton bound states in this case.  In those points $
\frac{\beta^{2}}{8 \pi} = \frac{2}{2n+ 3}\;,$ $ n= 1,2, \cdots$ we
consider, however, this is not the case by the following reason.

 Let the mass of a bound state be $m_{b}= 2m \sin w \;,$ $\; 0 < w <
\frac{\pi}{2}\;$.  If it appears in the direct channel, we write
  $ m_{b}^{2} = 2m^{2} + 2m^{2} \cos u $ , from which
\beqn
  u = \pi - 2w \;\;\;.
\eeqn
  From the assumed spectrum of eq.~(\ref{eq:spectrum}), the location
of the pole due to the $\ell$-th breather is
\beqn
  iu = i u_{\ell} = i \pi ( 1- \frac{2\ell}{2n +1}) \;\;\;.
\eeqn
  The corresponding pole in the crossed channel is at $ i ( \pi
-u_{\ell} ) $.  The mass of the would-be bound state is
\beqn
 m_{b^{\prime}} = 2m \sin \left( \frac{ \left( \pi - (\pi - u_{\ell} )
\right )}{2} \right) = 2m \sin \frac{\pi ( n- \ell + \frac{1}{2}
)}{2n+1} \;\;\;.
\eeqn
This is not the mass spectrum we assumed originally in
eq.~(\ref{eq:spectrum}) and the presence of the crossed channel pole
is simply contradictory to the rest of our discussion.

\section{Acknowledgements}
    We acknowledge Professor E. Date for valuable remarks on quantum
groups.  We also thank Katsushi Ito, Yutaka Matsuo and Hisao Suzuki
for useful discussions.

\newpage

\appendix

\section{A}

   $ U_{q}(\hat{s \ell}(2))$ is a universal enveloping algebra defined
by the elements $\{ \{ h_{i}, e_{i}, f_{i}\; \;i= 0,1 \}\}$ and the
following relations
\beqn
    \label{eq:hef} \left[ h_{i}, e_{j} \right] = a_{ij} e_{j}\;\;\;,
\;\;\; \left[ h_{i}, f_{j} \right] = - a_{ij} f_{j}\;\;\;, \;\;\;
\left[ e_{i}, f_{j} \right] = \delta_{ij} \frac{ q^{h_{i}} -
q^{-h_{i}} } { q - q^{-1}} \;\;\;,\;\;\; \\ e_{i}^{3} e_{j} -(q^{2}
+1+ q^{-2}) e_{i}^{2} e_{j} e_{i} + (q^{2} +1+ q^{-2}) e_{i} e_{j}
e_{i}^{2} - e_{i} e_{j}^{3} =0\;\;\;, \;\;{\rm i \neq j}\;\;,
\label{eq:qse} \\ f_{i}^{3} f_{j} -(q^{2} +1+ q^{-2}) f_{i}^{2} f_{j}
f_{i} + (q^{2} +1+ q^{-2}) f_{i} f_{j} f_{i}^{2} - f_{i} f_{j}^{3}
=0\;\;\;, \;\;{\rm i \neq j}\;\;, \label{eq:qsf}
\eeqn
  Here $a_{ij}$ is the generalized Cartan matrix for $\hat{s
\ell}(2)$~:
\beqn
   a_{ij} = \left( \begin{array}{cc} 2 & -2 \\ -2& 2 \end{array}
\right) \;\;\;.
\eeqn

 The co-product is
\beqn
\label{eq:cop}
  \Delta \left( q^{h_{i}/2} \right) &=& q^{h_{i}/2} \otimes
q^{h_{i}/2} \;\;\; \\ \Delta \left( e_{i} \right) &=& e_{i} \otimes
q^{-h_{i}/2} + q^{h_{i}/2} \otimes e_{i} \;\;\;, \\ \Delta \left
( f_{i} \right) &=& f_{i} \otimes q^{-h_{i}/2} + q^{h_{i}/2} \otimes
f_{i} \;\;\;.
\eeqn

  When $q$ is an $\ell$-th root of unity, one can restrict the algebra
by $e_{i}^{\ell} =0,\;\; f_{i}^{\ell} =0$ \cite{Lustig}.

  The following homomorphism has been used \cite{Jimbo} in
systematically constructing the trigonometric solutions of the
Yang-Baxter equation through eq.~(\ref{eq:rdelta}):
\beqn
\label{eq:hom}
  \phi (e_{0}) &=& \lambda f_{1} \;\;,\;\;\;\; \phi ( e_{1}) = e_{1}
\;\;\;, \nonumber \\ \phi (f_{0}) &=& \lambda^{-1} e_{1} \;\;,\;\;
\phi ( f_{1}) = f_{1} \;\;\;.
\eeqn
  Here $\lambda$ is a nonvanishing complex number which we identify to
be $ ie^{(2n+1) v}$ in the text.

\section{B}

  Let us summarize here some formulas for TBA in the case of diagonal
scattering.  Assume that the system is in a box with length $L$.  We
consider the scattering of ${\cal N}$ particles in which ${\cal
N}_{a}$ are species $a ~(a = 1\sim s)$.  Let $S_{ab}$ and $m_{a}$ be
the two-body S-matrix and the mass of the particle $a$ respectively.
Taking the logarithm of the equation of the periodic boundary
condition, we obtain
 \beq Lm_{a}{\rm sinh}v_{a} -i\sum_{j (\neq
i)=1}^{ {\cal N} } \log S_{ab}(v_{i}-v_{j}) = 2\pi n_{i},\ \ n_{i} \in
{\bf Z}\;\;\;.  \label{eq:Lg} \eeq
 Given a set of integers $ \{\{ n_{i} \}\} $, eq.~(\ref{eq:Lg}) determines the
location of the
rapidities $\{\{ v_{i} \}\} $.  The set of integers $
n_{1}^{(1)},\cdots,n_{{\cal N}_{1}}^{(1)},n_{{\cal
N}_{1}+1}^{(2)},\cdots, n_{{\cal N}_{1}+ {\cal N}_{2}}^{(2)}, n_{{\cal
N}_{1}+ {\cal N}_{2}+1}^{(3)},\cdots,n_{ {\cal N} }^{(n)}$ is such
that the subset $\{\{ n_{i}^{(a)} \}\}$ is ordered
$n_{i}^{(a)}<n_{i+1}^{(a)}$ as $v_{i}<v_{i+1}$.

 The density of states per unit length and unit rapidity $P_{a}$ and
the density of particles $\rho_{a}$ are respectively defined by \beq
P_{a}(v_{i}) = \lim
\frac{n_{i+1}^{(a)}-n_{i}^{(a)}}{L(v_{i+1}-v_{i})}\;\;, \;\;\;
\rho_{a}(v_{i})= \lim \frac{1}{L(v_{i+1}-v_{i})}\;\;\;.  \eeq Taking
the limit ${\cal N}, L \rightarrow \infty,{\cal N}/L=$ finite in eq.
(\ref{eq:Lg}), we obtain \beq \label{eq:dia} \frac{m_{a}}{2\pi}\cosh v
+ \sum_{b=1}^{s} \int_{-\infty}^{+\infty} \frac{dv '}{2\pi}
\phi_{ab}(v -v ')\rho_{b}(v ')	 =P_{a}(v)\;\;\;. \label{eq:Rho} \eeq
Here $\phi_{ab}$ is given by \beq \label{eq:phi} \phi_{ab}(v )=-i
\frac{d}{dv } {\rm log}S_{ab}(v )\;\;\;.  \eeq

Introducing $\epsilon_{a}(v)$ by \beq \frac{\rho_{a}(v )}{P_{a}(v )}
\equiv \frac{e^{ \frac{ \mu_{a}}{T} - \epsilon_{a}(v )}} {1+e^{ \frac
{ \mu_{a}}{T} - \epsilon_{a}(v )}}\;\;.  \label{eq:defep} \eeq The
free energy per unit length is given by \beqn \lim_{L \rightarrow
\infty} - \frac{1}{L} \log{\rm Tr}(e^{ \frac{ \sum_{a}\mu_{a}{\cal
N}_{a} }{T} }e^{- \frac{H}{T} })= \sum_{a=1}^{s} m_{a} \int
\frac{dv}{2\pi}{\rm cosh}v \log(1+e^{ \frac
{ \mu_{a}}{T}-\epsilon_{a}(v )}) \;\;\;, \eeqn where the
$\epsilon_{a}(v )$ are obtained as the solutions to \beq
\epsilon_{a}(v )= \frac{m_{a}}{T} {\rm cosh}v - \sum_{b (\neq a)
=1}^{s} \int \frac{dv '}{2\pi}\phi_{ba}(v '-v) \log(1+e^{ \frac
{ \mu_{b}}{T}-\epsilon_{b}(v ')}) \;\;\;. \label{eq:TBAeq} \eeq These
are the thermodynamic Bethe ansatz integral equations with chemical
potentials.

The entropy is given as a functional of $\rho_{a}(v)$ and
$\epsilon_{a}(v)$, \beq
\label{eq:ent}
     S=\sum_{a=1}^{n}\int_{-\infty}^{\infty} dv\rho_{a}
\left[(1+e^{\epsilon_{a}- \frac{\mu_{a}}{T} })
\log(1+e^{\epsilon_{a}- \frac{\mu_{a}}{T} })- (\epsilon_{a}- \frac
{ \mu_{a}}{T} )e^{\epsilon_{a}- \frac{\mu_{a}}{T} }	 \right]
\;\;\;.  \eeq

\section{C}
\label{sec:Rdil}

  The Rogers dilogarithm is defined by the integral \beq
\label{eq:dilog} L(x)=-\frac{1}{2} \int_{0}^{x}(\frac{{\rm log}\
y}{1-y} +\frac{{\rm log}(1-y)}{y})dy\;\;\; \nonumber \eeq for $x\in
{\bf C} \backslash ((-\infty,0) \cup (1,\infty))$\cite{RD,RD2}.
$$L(1)=\frac{\pi^{2}}{6},\ \ L(0)=0,\ \
L(\frac{1}{2})=\frac{\pi^{2}}{12}\;\;,\ \
L(\frac{\sqrt{5}-1}{2})=\frac{\pi^{2}}{10}\;\;,\ \
L(\frac{3-\sqrt{5}}{2})=\frac{\pi^{2}}{15}\;\;\;$$ are the only known
cases for which $L(x)$ can be evaluated explicitly.
($\frac{\sqrt{5}-1}{2}$ is an inverse of the golden ratio.)  Some
useful properties of Rogers dilogarithm are given by the following
equations: \beq L(x)+L(1-x)=L(1)=\frac{\pi^{2}}{6}\;\;\;,
\label{eq:L1} \eeq \beq L(\bar{x})=\overline{L(x)} \;\;\;, \eeq \beq
L(x)+L\left[\frac{1}{x}\right]= \frac{\pi^{2}}{3}+i\frac{\pi}{2}sgn
( \Im x)\log x \;\;\;.  \eeq For $x\in(0,1)$, $k$ integer and $q\equiv
e^{\frac{2\pi i}{2k+1}}$, \beq
L(x^{2k+1})=(2k+1)\sum_{j=0}^{2k}L(q^{j}x)+ 2\pi \sum_{j=1}^{k}j \arg
(1-q^{j}x), \eeq where $ \arg z$ is the argument of $z$, ${\bf C}
\backslash (-\infty,0]\rightarrow (-\pi,\pi)$.  \beq
L(a)+L(b)+L(c)+L(d)+L(e)=\frac{\pi^{2}}{2}\;\;, \label{eq:lwa} \eeq
where (a,b,c,d,e) is a 5-cycle, i.e.  $$a=1-cd\;,\ \ b=1-de\;,\ \
c=1-ea\;,\ \ d=1-ab,\ \ e=1-bc\;\;.$$ From eq.~(\ref{eq:lwa}), we
obtain \beqn L(x)+L(y) &=& L(xy)+L\left[\frac{x(1-y)}{1-xy}\right]+
L\left[\frac{y(1-x)}{1-xy}\right]\;\;\;, \\
L\left[\frac{1}{x^{2}}\right] &=& 2\left(L\left[\frac{1}{x}\right]-
L\left[\frac{1}{x+1}\right]\right)\;\;. \label{eq:Lm} \eeqn A very
useful summation property is \beq \sum_{k=2}^{n-2}
L\left[\frac{\sin^{2}\frac{\pi}{n}}{\sin^{2}\frac{k\pi}{n}}\right]=
\frac{2(n-3)}{n}L(1)\;\;\;. \label{eq:Lsum} \eeq

\section{D}

  Let us take the massless limit of eq.~(\ref{eq:epseq}) in the
presence of the purely imaginary chemical potentials.  We choose the
fermion number $f = \frac{1}{2}$ as an illustration.  The fermion
number of the state which has the eigenvalue given in
eq.~(\ref{eq:lamd2}) is $\frac{1}{2}n^{\prime} -\frac{1}{2}( {\cal N}-
n^{\prime}) =n^{\prime} -\frac{1}{2} {\cal N}$.  In the infinite
volume limit, we have \beqn \lim_{ L \rightarrow \infty}
\frac{n^{\prime}}{L} &=& \int dv \left( \sum_{k \in {\cal L} } \left
( P^{(k)}(v)- \rho_{+}^{(k)}(v) \right) + \sum_{\bar{k} \in
\overline{\cal L} } \rho_{+}^{(\bar{k})}(v) \right) \;\;\, \\ \lim_{ L
\rightarrow \infty} \frac{ {\cal N} }{L} &=& \int dv\rho_{\bf m} (v)
\;\;\;.  \nonumber \eeqn From eq.~(\ref{eq:mlpbc}), we find \beq \int
dv P^{(\ell)} (v) =(-1)^{\ell} \frac{1}{2}\int dv \rho_{\bf m} (v)
\;\;\;, \;\;\; \ell \in {\cal L} \oplus \overline{\cal L} \;\;\;.
\eeq Using this relation, we find the fermion number per unit length
${\cal F}$ to be \beq {\cal F} 	 =\int dv\sum_{\bar{k}\in
\overline{\cal L} } \rho_{+}^{(\bar{k})}- \sum_{k \in {\cal L} }
\rho_{+}^{(k)} \;\;\;.  \eeq This tells us that we must choose
\beqn
\label{eq:mu}
 \frac{\mu_{k}}{T} &=& - \frac{\mu_{\bar{k}} }{T} = -i\alpha \;\;\;,
\nonumber \\ {\rm and}\;\;\; \mu_{a} &=& \mu_{ {\bf m} }= 0 \;\;\;.
\eeqn

  Let $x_{\kappa}(\alpha) \equiv e^{-\epsilon_{\kappa}(0;\alpha)}$. With
eq.~(\ref{eq:mu}),  we find
  \begin{eqnarray}
\label{eq:algeq}
     \frac{1}{x_{a}}&=&\prod_{b=1}^{a-1}(1+x_{b})^{2b}
     (1+x_{a})^{2a-1}\prod_{c=a+1}^{s}(1+x_{c})^{2a}
     (1+x_{  {\bf m} })^{a}  \;\;\;,                             \nonumber \\
     \frac{1}{x_{{\bf m} }}&=&  \left( \prod_{b=1}^{n}(1+x_{b})^{b} \right)
     (1+x_{{\bf m} })^{n}(1+2x_{even}\cos\alpha+x_{even}^{2})^{-\frac{n+1}{2}}
\nonumber \\
    &\times& (1+2x_{odd}\cos\alpha+
  x_{odd}^{2})^{\frac{n}{2}}\;\;\;.       \nonumber \\
     \frac{1}{x_{even}}   &=&  x_{odd}  =
     (1+x_{ {\bf m} })^{-\frac{1}{2}}  \;\;\;.
  \end{eqnarray}
  By try and error, we find a solution  to eq.~(\ref{eq:algeq})
  \begin{eqnarray}
     e^{\epsilon_{a}(0;\alpha)}  &=&
     \frac{\sin\frac{a\alpha}{2n+3}\sin\frac{(a+2)\alpha}{2n+3}}
          {\sin^{2}\frac{\alpha}{2n+3}}  \;\;\;, \;\;\;
     e^{\epsilon_{m}(0;\alpha)} =
     \frac{\sin^{2}\frac{(n+1)\alpha}{2n+3}}
          {\sin\alpha \sin\frac{\alpha}{2n+3}} \;\;\;, \;\;\;  \nonumber \\
     e^{\epsilon_{even}(0;\alpha)} &=&  e^{-\epsilon_{odd}(0;\alpha)} =
     \frac{\sin\frac{(n+1)\alpha}{2n+3}}
          {\sin\frac{(n+2)\alpha}{2n+3}}  \;\;\;.
  \end{eqnarray}
  As for $\epsilon_{\lambda}(\infty;\alpha)$, the algebraic equations are
  the same   as
the $\alpha=0$ case. So are the solutions.
   Thus
\beqn
 \epsilon_{\lambda}(\infty;\alpha) =
  \epsilon_{\lambda}(\infty; 0) \equiv  \epsilon_{\lambda}(\infty)\;\;\;.
\eeqn

\end{document}